\documentclass[reprint, amsmath,amssymb,aps,prl,superscriptaddress,longbibliography,]{revtex4-2}

\usepackage [english]{babel}
\usepackage{graphicx}					% Include figure files
\usepackage{dcolumn}					% Align table columns on decimal point
\usepackage{bm}							% bold math
\usepackage[colorlinks=true,linkcolor={blue},pdfstartview=Fit,citecolor={blue}]{hyperref}

\usepackage{mathtools}
\usepackage{physics}

\usepackage[dvipsnames]{xcolor} 
\usepackage[normalem]{ulem}

\DeclareMathOperator {\e}{e}				% e for exponential should be roman

\renewcommand{\thesection}{\arabic{section}}

\newcommand{\beginsupplement}{%
        \setcounter{table}{0}
        \renewcommand{\thetable}{S\arabic{table}}%
        \setcounter{figure}{0}
        \renewcommand{\thefigure}{S\arabic{figure}}%
        \setcounter{section}{0}
        \renewcommand{\thesection}{S\arabic{section}}%
     }

\begin{document}

\title{Anomalous heating in a colloidal system}

\author{Avinash Kumar}
\affiliation{Department of Physics, Simon Fraser University, Burnaby, British Columbia V5A 1S6, Canada}
\author{Rapha\"el Ch{\'e}trite}%
\affiliation{%
 Laboratoire J A Dieudonn{\'e}, UMR CNRS 7351
 Universit{\'e} de Nice Sophia Antipolis, Nice, France}%
\author{John Bechhoefer}
\email{johnb@sfu.ca}
\affiliation{Department of Physics, Simon Fraser University, Burnaby, British Columbia V5A 1S6, Canada}

\date{\today}

\begin{abstract}
We report anomalous heating in a colloidal system, the first observation of the  inverse Mpemba effect, where an initially cold system heats up faster than an identical warm system coupled to the same thermal bath. For an overdamped, Brownian colloidal particle moving in a tilted double-well potential, we find a non-monotonic dependence of the heating times on the initial temperature of the system, as predicted by an eigenfunction expansion of the associated Fokker-Planck equation. By carefully tuning parameters, we also observe a ``strong'' version of anomalous heating, where a cold system heats up exponentially faster than systems prepared under slightly different conditions. 

\end{abstract}


\maketitle

\textit{Introduction.}---Can an initially cold system heat up faster than an initially warm system that is otherwise nominally identical?  Naively, one would assume that a slowly heating object relaxes to the temperature of its surroundings exponentially, passing through all the intermediate temperatures. A system that is initially at a cold temperature should then take longer to heat than a system initially at a warm temperature. However, for rapid heating, a system may evolve towards equilibrium so that its intermediate states are not in thermal equilibrium with the surrounding heat bath and are not characterized by a unique temperature. In such cases, the possibility of anomalously fast heating has recently been predicted and confirmed in numerical studies of an Ising antiferromagnet \cite{lu2017nonequilibrium}.  Further numerical studies suggest that these effects may be seen in a wide variety of systems, including fluids with inelastic \cite{lasanta2017hotter,rajesh2020granularMaxwellgas,santos2020nonlineardrag} and elastic  \cite{takada2020mpemba,gomez2020mpemba} collisions and spin glasses \cite{baity2019mpemba}. 

Although anomalous heating is a recent prediction, an analogous anomaly for cooling and freezing has been noted in observations of water dating back to 350 BC \cite{Webster1923aristotle}.  Its first systematic study was done in 1969 by Mpemba and Osborne, who concluded that hot water could begin to freeze in a time shorter than that required for cold water \cite{mpemba1969cool}.  This phenomenon has since been dubbed the \textit{Mpemba effect} and was followed up further experiments on water \cite{Kell1969,woj1988,Auerbach1995supercooling,
esposito2008mpemba,katz2009hot,firth1971cooler,burridge2020observing}, accompanied by some controversy, tracing back to the difficulty of obtaining reproducible results \cite{burridge2016questioning,katz2017reply}.  Proposed mechanisms for the effect include evaporation \cite{Kell1969, vynnycky2010evaporative, mirabedin2017}, convection currents \cite{freeman1979cooler,vynnycky2012axisymmetric,vynnycky2015convection}, dissolved gases and solutes \cite{freeman1979cooler,woj1988,katz2009hot}, supercooling \cite{Auerbach1995supercooling,esposito2008mpemba}, and hydrogen bonds \cite{zhang2014hydrogen,Tao2016Hbonding}. 

In an effort to understand the Mpemba effect in more generic terms, Lu and Raz introduced a theoretical picture that related the effect to the geometry of system dynamics in a state space whose elements are defined by the amplitudes of eigenmodes of the system dynamics \cite{lu2017nonequilibrium}.  A fast quench can then lead a system to follow a nonequilibrium path through state space to equilibrium that is shorter than the path traced out by a slowly cooling system.  In recent work, we showed that this kind of Mpemba effect is present in a system consisting of a colloidal particle immersed in water and subject to a carefully designed potential \cite{kumar2020exponentially}.  From this point of view, the dynamics of cooling and heating obey similar principles, and anomalous heating represents an \textit{inverse} Mpemba effect. Yet, despite a formal similarity between the cases of heating and cooling \cite{lu2017nonequilibrium}, anomalous heating has not previously been seen experimentally. Indeed, as we shall see, subtle differences between high- and low-temperature limits generically make the inverse effect more difficult to observe experimentally. Moreover, the mechanism for the inverse effect does not depend on the presence of metastability, which played a crucial role in the forward case analyzed in Ref.~\cite{chetrite21}.

Here we present the first experimental evidence for the inverse Mpemba effect. Our results agree quantitatively with predictions based on the theoretical framework of Lu and Raz \cite{lu2017nonequilibrium}.   We also observe a \textit{strong} version  \cite{Marija2019robust} of the effect, where, for a carefully chosen initial temperature, a system heats up exponentially faster than systems that were initially at different temperatures.

\begin{figure}[ht!]
\centering
\includegraphics{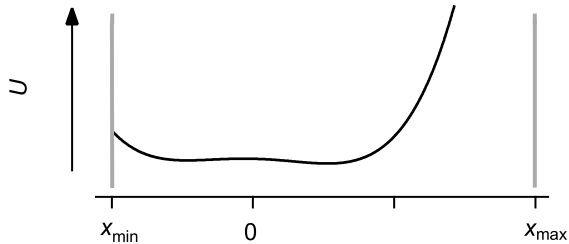}
\caption[Energy landscape of the inverse Mpemba effect] {
Schematic of the energy landscape $U(x)$ used to explore the inverse Mpemba effect, set asymmetrically ($\alpha=2$) within a box $[x_\text{min},x_\text{max}]$ with potential walls with finite slope at the domain boundaries. Because of the tilt in the potential, the positions of the left and right minima, and also the energy barrier, shift and are approximately at $-$37.2 nm, +42.2 nm, and $-$5.0 nm, respectively.}
\label{Fig:pot_ime} 
\end{figure}

\textit{Experimental setup.}---In our experiment, a Brownian particle (silica bead, \O 1.5 \textmu m) is subjected to forces exerted by an external potential. The potential is a one-dimensional double well, created by a feedback trap based on optical tweezers \cite{kumar2018nanoscale,albay2018optical}. We place the potential asymmetrically in the domain $[x_\text{min},x_\text{max}]$ as
\begin{equation}
U(x) \equiv 
\begin{dcases}
   -F_\text{max} x \qquad & \phantom{MNI} x < x_\text{min} \\[1.5ex]
   \phantom{-} U_0(x)     \qquad      & \phantom{M}  x_\text{min} \leq  x \leq x_\text{max}\\[1.5ex]
   \phantom{-} F_\text{max} x  \qquad & \phantom{MNI}  x > x_\text{max}\,, \\[1.5ex]
\end{dcases}
\label{Eq:pieceDW_IME}
\end{equation}
where $U_0(x)$ is given by
\begin{equation}
	U_0(x) = E_\text{b} \left( (1-x^2)^2  - \frac{1}{2}x \right) \,,
\label{Eq:U_IME_fin}  
\end{equation}
with a very low barrier $E_\text{b} = 0.0002 \, k_\textrm{B}T_\textrm{b}$, with $k_\textrm{B}$ the Boltzmann constant and $T_\textrm{b}$ the bath temperature.  The position $x$ is measured in units of $x_\text{m} = 40$ nm (Fig.~\ref{Fig:pot_ime}). The geometric asymmetry in the potential is defined by the parameter $\alpha = |x_\text{max}/x_\text{min}|$.

Our setup has steep walls at the domain boundaries corresponding to the maximum force $F_\text{max} \approx 20$ pN $\approx 5~k_\textrm{B}T_\textrm{b}$/nm applied by the optical tweezers \footnote{See Appendix for further discussion of the experimental setup and the Leidenfrost effect.}.  The nearly vertical walls confine particle motion to a box with $\alpha \approx 2$, in which a particle relaxes.  Although we have defined $\alpha$ using the geometric size of the domains measured with respect to the origin, the barrier position is shifted to the left of the origin by $\approx 5$ nm due to the tilt in the potential. This offset results in a bias of $\approx 2 \%$ in $\alpha$ and $\approx 3\%$ shift in the equilibrium probabilities for the left and the right states.

\textit{Quenching protocol.}---An instantaneous ``heating quench'' in our experiments is a three-step process: (i) prepare the initial state of the system corresponding to the Boltzmann distribution $\pi(x;T_0) \propto \text{exp}[-U(x)/k_\text{B}T_0]$ at an initial temperature $T_0$; (ii) release a particle at a position sampled from the initial distribution $\pi(x;T_0)$; and (iii) record the trajectories of the particle as it relaxes in a bath at temperature $T_\text{b}$. The initial positions are sampled assuming $U(x)$ to have infinite potential walls at the domain boundaries. Once the particle is released into the bath, it is always at the bath temperature. We repeat the quenching protocol $N=5000$ times, with each cycle $60$ ms long, to create a statistical ensemble of the state of the system at each time step $\Delta t = 10$ \textmu s.  The dynamics of the particle after the quench in the potential $U(x)$ can be described by the overdamped Langevin equation
\begin{align}
	\dot{x} = -\frac{1}{\gamma}U'(x) + \sqrt{\frac{2k_\text{B}T_\text{b}}{\gamma}}\eta(t) \,,
\label{Eq: Langevin-eq}
\end{align}
where $\gamma$ is the Stokes friction coefficient and $\eta$ Gaussian white noise, with $\langle \eta(t) \rangle = 0$ and $\langle \eta(t) \, \eta(t')\rangle = \delta(t-t')$.

Although the initial and final states in our experiment obey Boltzmann distributions at temperatures $T_0$ and $T_\text{b}$, the intermediate states $p(x,t)$ are not in equilibrium. The intermediate state typically does not have the form of a Boltzmann distribution for any temperature $T$. For this reason, instead of trying to define an intermediate effective temperature, we measure the \textit{distance} $\mathcal{D}$ between the intermediate state $p(x,t)$ and the equilibrium state $\pi(x;T_\text{b})$ \cite{lu2017nonequilibrium,kumar2020exponentially}.  From equivalent alternatives \cite{lu2017nonequilibrium}, we choose the $L_1$ distance \cite{cover06} for the analysis of particle trajectories in our experiments. This distance is defined as the absolute difference between $p_i$ and $\pi_i$,
\begin{align}
	\mathcal{D}[p(x,t);\pi(x;T_\textrm{b})] \equiv \mathcal{D}(t) = \sum_{i=1}^{N_\text{b}}|p_i-\pi_i| \,.
\label{eq: L1-dist}
\end{align}
Here $p_i \equiv p(x_i,t)$ is the frequency estimate of the probability for a measured position $x$ at a time $t$   in the interval $[x_i,x_{i+1})$, where $x_i \equiv x_\text{min} + (i-1)\Delta x$, with $\Delta x = (x_\text{max}-x_\text{min})/N_\text{b}$ and $N_\text{b}$ the number of bins. Similarly, $\pi_i$ is the frequency estimate of the Boltzmann distribution at $T_\text{b}$.

\textit{Inverse Mpemba effect in an asymmetric potential.}---
To determine how the inverse Mpemba effect depends on the initial temperature of the system, we release the particle in a bath of fixed temperature (Fig.~\ref{Eq:pieceDW_IME}). 
\begin{figure}[ht!]
\centering
 \includegraphics{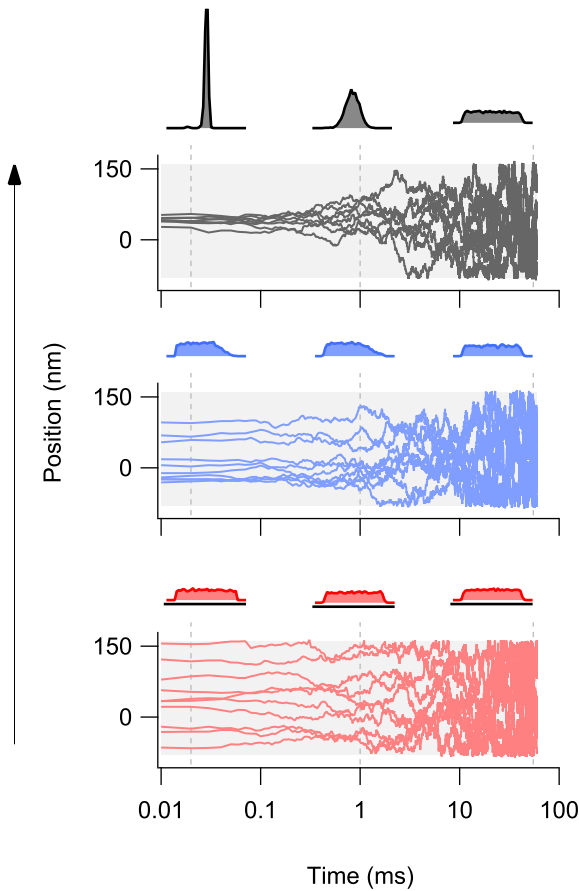}
\caption{Dynamic trajectories relaxing to equilibrium at a hot temperature. Ten trajectories of a particle released from the equilibrium distribution at temperatures $T_0 = 4\times10^{-4}T_\text{b}$ (black), $4\times10^{-3}T_\text{b}$ (blue) and $T_\text{b} = 1$ (red) into the hot bath, with the evolving probability density $p(x,t)$ shown for three times (estimated based on 5000 trajectories) on a logarithmic time scale. The shaded gray region corresponds to a box size of $x_\text{max}-x_\text{min} = 240$ nm.}
\label{Fig:traj_IME}
\end{figure}
After a particle is released in the bath at $t = 0$ at a low temperature $T_0$, it moves stochastically in response to thermal fluctuations and potential-gradient forces and finally equilibrates with the bath, which is at temperature $T_\text{b}~>~T_0$. 

\begin{figure}[b!]
\centering \includegraphics[width=3.3in]{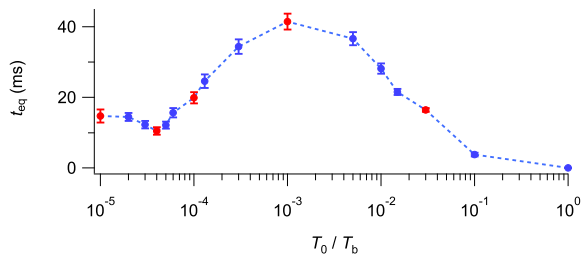}
\caption{Equilibration times for systems at different initial temperatures.  Red markers denote initial temperature points whose $\mathcal{D}(t)$ dynamics are displayed in Fig.~\ref{Fig:dist_ime_fin}.}
\label{Fig:teq_ime_finite}
\end{figure}

Figure~\ref{Fig:traj_IME}(a)--(c) shows example time traces of evolution in the potential $U(x)$. Figure~\ref{Fig:teq_ime_finite} shows the measured times to reach equilibrium for systems that start at different initial temperatures. As the initial temperature of the system decreases from $T_0/T_{\text{b}}=1$ to $\approx 10^{-3}$, the equilibration time increases monotonically and follows normal heating ($t_\text{c}>t_\text{w}$). However, for the lower initial temperature range $10^{-3}>T_0/T_{\text{b}}>10^{-5}$, the equilibration time decreases as the initial state of the system gets colder. Such a behavior corresponds to anomalous heating where a cold system takes less time to heat up than a warm system, i.e., $t_\text{c}<t_\text{w}$. For lower temperatures ($T_0/T_{\text{b}}<1\times 10^{-5}$), the equilibration time increases again, exhibiting normal heating. Thus, we observe a sequence of normal, anomalous, and normal regimes for relaxation to thermal equilibrium.

\textit{Analysis in the high-temperature limit.}---In Eq.~\ref{Eq:U_IME_fin}, the variations in $U_0(x)$ throughout the domain $[x_\textrm{min},x_\textrm{max}]$ are $\ll k_\text{B}T_\text{b}$, implying that dynamics at the bath temperature approximate ordinary diffusion. To simplify the analysis of the relaxation trajectories in $U(x)$ at a finite temperature $T_\text{b}$, we model the bath as being at an effectively infinite temperature with no energy barrier.  Further, we approximate the walls $\pm F_\textrm{max}x$ as being infinitely steep.  The particle then freely diffuses in a domain with walls at $x_\textrm{min}$ and $x_\textrm{max}$.  

Approximating the bath as being at a very high temperature and the walls as infinitely steep simplifies the analysis in three ways:  (i) the equilibrium state is a uniform distribution; (ii) the Fokker-Planck operator is self-adjoint, so that left and right eigenfunctions are identical; (iii) the eigenfunctions have simple analytic expressions.

\begin{figure}[hb!]
\centering
 \includegraphics{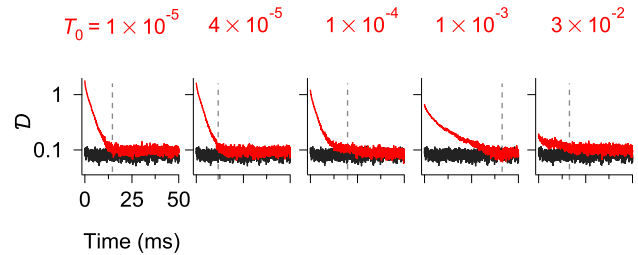}
\caption{$L_1$ distances $\mathcal{D}(t)$ for systems that heat up in a bath at temperature $T_\text{b}=1$, starting at  the temperatures $T_0$ indicated above each graph.  The thin vertical lines indicate the times when systems first reach thermal equilibrium (within noise levels).}
\label{Fig:dist_ime_fin} 
\end{figure}

In this high-temperature approximation, the Fokker-Planck equation describing the probability density $p(x,t)$ of particle positions reduces to the heat equation,
\begin{align}
	\pdv{p}{t}&=  \left[-\frac{1}{\gamma}\pdv{}{x}U'(x) + \frac{k_\text{B}T_\text{b}}{\gamma} \pdv[2]{}{x}\right] p(x,t)\nonumber \\
	&\approx\frac{k_\text{B}T_\text{b}}{\gamma} \pdv[2]{p}{x}\,
	\equiv \mathcal{L}_\text{free}\,p \,,
\label{eq:FP_IME}
\end{align}
subject to no-flux boundary conditions at $x~=~\{ x_\textrm{min}, x_\textrm{max} \}$.  Note that this high-temperature limit is complementary to but less familiar than the low-temperature limit, which leads to metastability phenomena~\cite{Freidlin12}. Indeed, the high-noise limit of the Langevin equation has recently stimulated wide interest because of its relation to the strong-measurement limit of quantum measurements \cite{bauer2018,bernardin20}. 

In the heat-equation limit, the probability density $p(x,t)$ can be written as an infinite sum of eigenfunctions $v_k$ of $\mathcal{L}_\text{free}$ with associated eigenvalues
\begin{align}
	-\lambda_k  = \frac{k_\text{B}T_\text{b}}{\gamma} \frac{\pi^2 (k-1)^2}
		{(x_\text{max}-x_\text{min})^2} \,,
\label{Eq: eig-vals}
\end{align}
ordered so that $0=\lambda_1 < \lambda_2 < \cdots$. At large but finite times, we assume that the contribution of the eigenfunctions $v_k(x,T_\text{b})$ decreases exponentially for $k>2$. Thus, the probability density can be approximated by 
\begin{align}
	p(x,t) ~\approx~ \pi(x;T_\text{b}) + \underbrace{a_2(T_0) 
	\e^{-\lambda_2 t}}_{a_2(t)}v_2(x; \alpha,T_\text{b}) \,,
\label{eq:p2_IME}
\end{align}
where the coefficient $a_2$ depends on the initial temperature $T_0$, as well as on the bath temperature $T_\textrm{b}$ and $a_2(t)$ represents the dynamics of the $v_2$ mode amplitude during thermalization. Generally, $a_2$ is a measure of the overlap between the second left eigenfunction and the initial state of the system \cite{kumar2020exponentially},
\begin{align}
	a_2(T_0) = \langle u_2(x;T_\text{b}) | \pi(x;T_0) \rangle \,.
\label{Eq: a2}
\end{align}

In the high-temperature limit, the spatial eigenfunctions of the diffusion equation are \cite{arfken11} 
\begin{align}
	u_k = v_k = \frac{1}{Z'}\,\text{cos}\left[(k-1) \pi \left(\frac {x-x_\text{min}}
		{x_\text{max}-x_\text{min}}\right)\right] \,,
\label{Eq:eig_ime}
\end{align} 
where $Z'$ is the normalization constant, defined such that $\langle u_k|v_k \rangle = 1$ with $k = 1,2,\cdots$.  Note that left and right eigenfunctions are identical for the diffusion equation but usually differ for the Fokker-Planck equation associated with non-zero potentials \cite{risken89}. 

Since anomalous heating (inverse Mpemba) is associated with $a_2$ coefficients where $|a_2(T_\text{w})|>|a_2(T_\text{c})|$, a non-monotonic temperature dependence of the $a_2$ coefficients leads to anomalous heating effects.  But these coefficients are not directly accessible in experiments. Instead \cite{kumar2020exponentially}, we measure a quantity $\Delta \mathcal{D} \propto |a_2|$ as a function of initial temperature from $\mathcal{D}(t)$, defined in Eq.~\ref{eq: L1-dist}. 

\begin{figure}[h!]
\centering
 \includegraphics[width=3.3in]{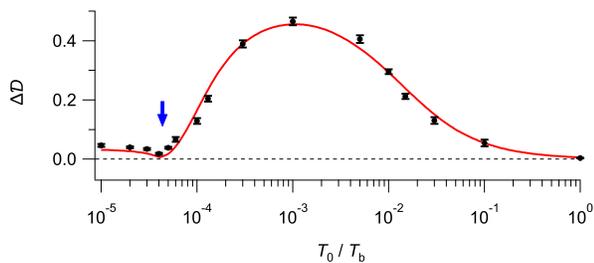}
\caption{Measurements of decay amplitude $\Delta \mathcal{D}$ for different initial temperatures $T_0$. Markers denote experimental measurements, and the solid red line is based on the  $|\Delta \mathcal{D}|$ values calculated from the FPE in the high-temperature limit.  The arrow indicates the temperature at which the strong inverse Mpemba effect occurs.  Error bars represent one standard deviation and are calculated from the fits.}
\label{Fig:delD_ime}
\end{figure}

Figure~\ref{Fig:delD_ime} shows the non-monotonic temperature dependence of $\Delta \mathcal{D}$. The $\Delta \mathcal{D}$ values correlate with the measured equilibration times. To see the agreement of the measured values of $\Delta \mathcal{D}$ for the potential at finite temperature with theoretical predictions based on the potential at a high temperature, we explicitly calculate $a_2$ coefficients using Eqs.~\ref{Eq: a2} and \ref{Eq:eig_ime}. We fit the data to a single parameter, an overall proportionality constant. The fit leads to $1.48 \pm 0.03$, which agrees to $\approx 5\%$ with the calculated value, $\approx 1.56$.

At initial temperature $T_0= 4 \times 10^{-5}$, where $a_2(T_0) \approx 0$ (Fig.~\ref{Fig:delD_ime}, blue arrow), the decay is dominated by $\lambda_3$ and represents an exponential speed-up of the heating process compared to decays at temperatures where $a_2(T_0)~\neq~0$. Such a situation corresponds to the \textit{strong} inverse Mpemba effect \cite{klich2018solution}. The transient decay at the time scale set by the eigenvalue $\lambda_2^{-1}\approx 16.66$~ms disappears, and the system decays instead at a rate $\lambda_3^{-1} \approx 4.15$ ms.  In summary, for $|a_2(\alpha,T_\text{w})|>|a_2(\alpha,T_\text{c})|$, the initially warm system lags the initially cold system, and the inverse Mpemba effect is observed.

\textit{Discussion.}---Our results give clear experimental evidence for the inverse Mpemba effect in a simple setup. The non-monotonic dependence of the equilibration time on the initial temperature of the system can be understood through the non-monotonicity of $a_2$ coefficients. We observed the inverse Mpemba effect for a quench for the case of a heat bath whose average energy greatly exceeded the range of potential variation.  We used this feature to model system dynamics in a high-temperature limit where the relaxation dynamics are governed by a simple heat-diffusion equation.  Using analytic expressions for the eigenfunctions, we obtained the $a_2$ coefficients as a function of initial temperature.  We found evidence for the strong inverse Mpemba effect, special temperatures where the systems heat up exponentially faster than those at other initial temperatures.

Previous experiments on the forward Mpemba effect showed a clear separation of the time scales determined by the eigenvalues $\lambda_2$ and $\lambda_3$ \cite{kumar2020exponentially}. We can offer some insight as to why it is easier to observe the forward Mpemba effect than the inverse effect: When a system relaxes to a bath at temperature $T_\text{b}$, the time-scale separation between the decay curves corresponding to $\lambda_3$ and $\lambda_2$ depends on the ratio $\Lambda = \lambda_3/\lambda_2$. In particular, in order to measure the $\Delta \mathcal{D}$ values, one fits the part of the decay curve that corresponds to $\lambda_2$. Thus, the greater the value $\Lambda$, the easier the accurate measurement of the $\Delta \mathcal{D}$ values.

For the forward Mpemba effect studied in Ref.~\cite{kumar2020exponentially}, the system cools from a hot temperature to a cold temperature in a double-well potential, and the ratio $\Lambda$ of eigenvalues $\lambda_3$ to $\lambda_2$ (i.e., $\Lambda^{*}_\text{for} \equiv \lambda_3/\lambda_2$) is $\approx 16.1$. However, for the inverse Mpemba effect studied here, the ratio of eigenvalues is $\Lambda^{*}_\text{inv} \approx 4.0$. Thus, $\Lambda^{*}_\text{inv}$ is about four times smaller than $\Lambda^{*}_\text{for}$ in the case of heating, implying that the forward effect will be easier to observe than the inverse effect.  Indeed, our observations of the inverse effect required an ensemble of 5000 trajectories to obtain results that are statistically similar to results for the forward case obtained with only 1000 trajectories.

\begin{figure}[h!]
\centering
 \includegraphics[width=3.3in]{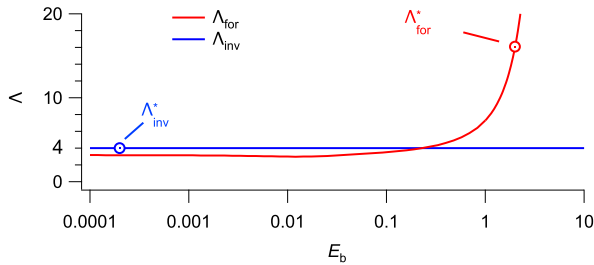}
\caption{Ratio $\Lambda$ of eigenvalues $\lambda_3$ to $\lambda_2$ of the Fokker-Planck operator as a function of barrier height $E_\text{b}$ at the bath temperature $T_\text{b} = 1$. The red curve is for the double-well potential used in Ref.~\cite{kumar2020exponentially} for the forward Mpemba experiments, and the blue curve is for the approximately flat potential used in the inverse Mpemba experiments. The hollow red  marker denotes the ratio $\Lambda^{*}_\text{for}$ used in the forward Mpemba experiments, that corresponds to $E_\text{b} \approx 2$. The hollow blue marker denotes the ratio $\Lambda^{*}_\text{inv}$ for the flat potential ($E_\text{b} \approx 0$) used in the inverse Mpemba experiments.}
\label{Fig:eigs_ratio}
\end{figure}

Are these general features of $\Lambda_\text{for}$ and $\Lambda_\text{inv}$ or are they special to our potential?  Since the dynamics of $a_2(t)$ correspond to hops over the barrier, we expect that the ratio of eigenvalues $\lambda_3$ to $\lambda_2$ depends on the barrier height $E_\text{b}$ as $\Lambda_\text{for} \sim \text{exp}[E_\text{b}/k_\text{B}T_\text{b}] \gg 1$, an intuition confirmed by a rigorous analysis in general \cite{berglund13,kolokoltsov2000} and by numerical solution of the Fokker-Planck equation for our potential in particular (Fig.~\ref{Fig:eigs_ratio}, red curve). However, for the high-temperature limit, Eq.~\ref{Eq: eig-vals} shows that $\Lambda_\text{inv} = 4$ always (Fig.~\ref{Fig:eigs_ratio}, blue curve).  Thus, the ratio of eigenvalues $\Lambda$ can be much higher in the forward case than in the reverse case, and, as a result, the forward effect is generically easier to observe experimentally than the inverse effect.

In this paper, we offer evidence for anomalous heating in a colloidal system, complementing the more familiar scenario for anomalous cooling.  Other \textit{memory-dependent} relaxation phenomena \cite{keim19} are worth exploring further.  For example, Gal and Raz show that an initial cooling quench followed by a heating quench can speed up heating times exponentially, even in systems that would not otherwise exhibit the inverse Mpemba effect \cite{gal20}.  In the \textit{Kovacs effect}, protocols that drive the system out of equilibrium can produce a non-monotonic relaxation, even after the forcing has ceased \cite{kovacs1964transition,kovacs1979isobaric,morgan20,militaru2021kovacs}.  Finally, in the \textit{Leidenfrost effect}, a water droplet placed on a hot surface survives evaporation longer than one placed on a warm surface. First described in the $18^{\text{th}}$ century \cite{leidenfrost1756aquae}, it shares some features with the inverse Mpemba effect (see Appendix, including Table S1).  In particular, it also involves a counterintuitive, non-monotonic temperature dependence of the time to reach the final state, and the nonequilibrium forcing is via heating.  It would be interesting to re-examine this well-known phenomenon from the perspectives developed here.

\begin{acknowledgments}
JB and AK acknowledge funding from Discovery and RTI grants from the National Sciences and Engineering Council of Canada (NSERC). RC acknowledges support from the Pacific Institute for Mathematical Sciences (PIMS), the French Centre National de la Recherche Scientifique (CNRS).
\end{acknowledgments}

\bibliography{references_IME}

\section{Appendix} 
\beginsupplement 

\begin{figure}[ht]
\centering
 \includegraphics[width=6in]{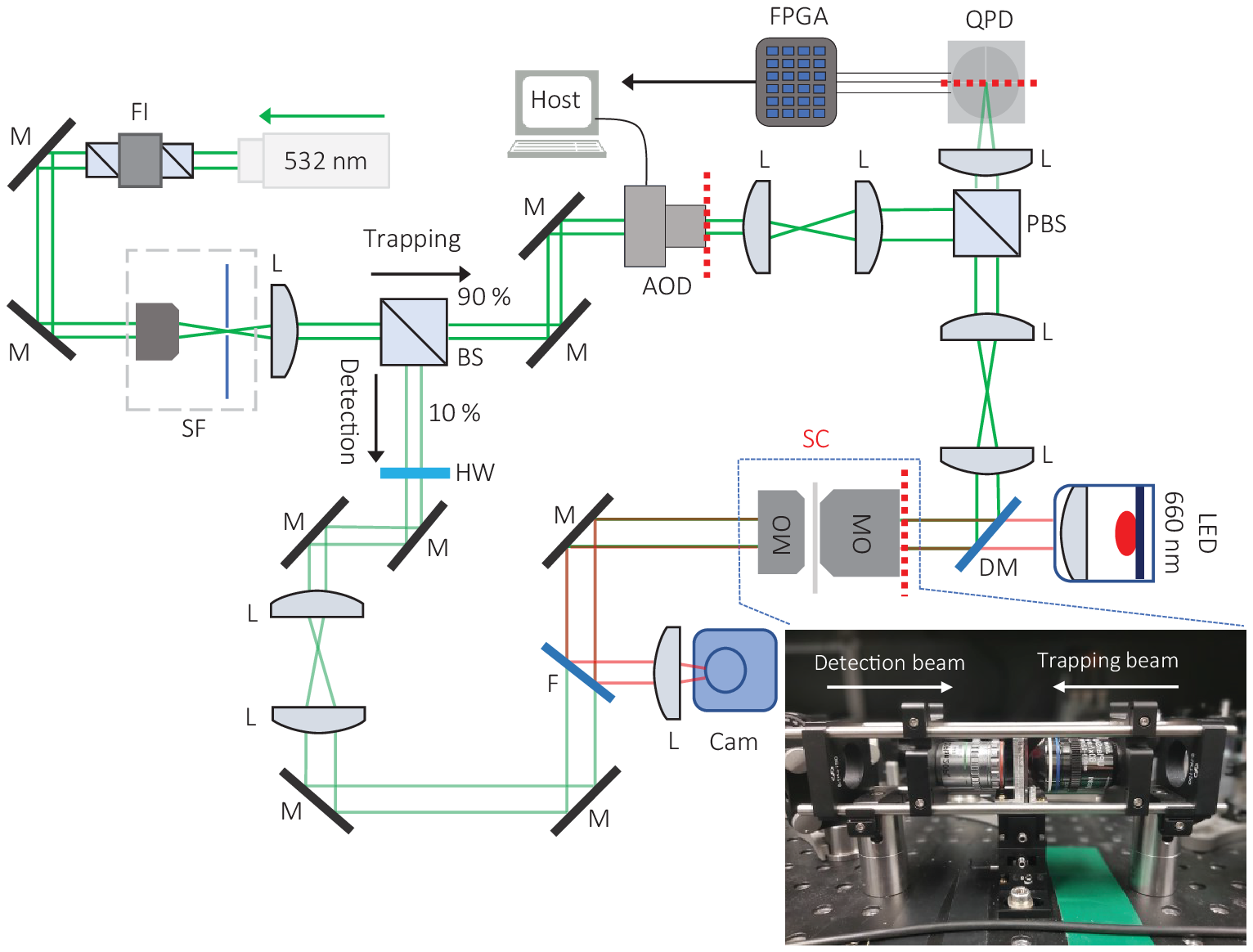}
\caption{Schematic of the feedback-trap setup. FI = Faraday Isolator, M = Mirror, SF = Spatial Filter, BS = Beam Splitter (non-polarizing), AOD = Acousto-Optic Deflector, L = Lens, MO = Microscope Objective, SC = Sample Chamber, PBS = Polarizing Beam Splitter, HW = Half-Wave Plate, F = Short-Pass Filter, QPD = Quadrant Photodiode, DM = Dichroic Mirror, PD = Photodiode, CS = Cover-Slip, Cam = Camera.
Planes conjugate to the back-focal plane of the trapping objective are shown in red-dashed
lines. An image of the cage-system consisting of the trapping and detection objectives is shown on the bottom-right corner.}
\label{Fig:setup}
\end{figure}

\textit{Feedback trap Setup.---}Our optical tweezers setup is built on a vibration isolation table supporting a home-built microscope.  We trap a colloidal particle (silica bead, \O 1.5 \textmu m, Bangs Laboratories). A linearly polarized 532 nm laser (Nd:YAG, Coherent Genesis MX STM-series, 1 Watt) is used for trapping and detection (Figure~\ref{Fig:setup}). In our experiment, we use acousto-optic deflectors (AODs) to steer the trap position in the trapping plain, placed at a plane conjugate to the back focal plane of the trapping objective. The details of the experimental setup are described in Ref.~\citep{kumar2018nanoscale}. A feedback scheme is used to create the virtual potentials used in the inverse Mpemba experiments \cite{kumar2018nanoscale,kumar2020exponentially}. In a feedback trap, one (1) observes the position of the particle, (2) calculates the force based on its position in the user-defined potential, and (3) applies that force in each loop at a deterministic time step of $\Delta t = 10$ \textmu s \cite{cohen05,gavrilov2014real}. In our experiment, the force is applied by moving the trap center relative to the bead position. The force generated by the displacement of the trap center is approximated as $F_n = -\beta x_{n-1}$, where $\beta = \Delta t/t_r$ is a proportional feedback constant where $t_r$ is the relaxation time of the underlying physical potential, and where $x_n$ is the particle position at time $t_n = n \Delta t$.

Compared to our previous results based on the newly developed feedback traps \cite{kumar2018optical,kumar2018nanoscale}, we have improved the mechanical stability of the setup by installing the trapping and detection objectives on a cage system (Figure~\ref{Fig:setup}). Thus, mechanical drifts due to temperature changes of the surroundings have reduced effects on the particle position. Nonetheless, we have drifts in the particle position on the order of 1 nm s$^{-1}$. We, thus, limit the heating cycle to 0.1 s and correct for the small drift before the next cycle.

%\section{Leidenfrost effect}

\textit{Leidenfrost effect.---}Although, to our knowledge, there has been no experimental evidence for the inverse Mpemba effect in any system, there does exist a well-known heating phenomenon, known as the \textit{Leidenfrost effect}, that dates back to the $18^{\text{th}}$ century \cite{leidenfrost1756aquae}. This phenomenon occurs when liquid droplets are deposited on hot solid surfaces, and a layer of vapor is formed between the droplet and substrate. The high-pressure vapor layer prevents contact between the hot surface and the droplet. The layer thus reduces the heat transfer between them, allowing the droplets to  survive much longer than normally expected. When the surface temperature is lower than the boiling point of the liquid, the droplets spread over the substrate to form a thin layer and evaporate slowly. Upon further increase in temperature, a maximum rate of evaporation is achieved at a critical temperature (also known as the  Nukiyama temperature, $T_\text{N}$) corresponding to the minimum survival time of the droplets \cite{nukiyama1966maximum}. Beyond $T_\text{N}$, the survival time rapidly increases and reaches a maximum value at the Leidenfrost point (LP) temperature. Thus, the effect is characterized by a significant reduction in heat transfer from a heated body to liquids when the temperature of the body belongs to a range of temperatures between $T_\text{N}$ and LP. In this range, the survival times of the liquid droplets increase with the temperature of the surface \cite{walker2010boiling,bernardin1999leidenfrost}. Beyond the LP, the survival time again decreases, indicating a non-monotonic temperature dependence of the survival time on the temperature of the surface. 

Although the effect is not the same as the inverse Mpemba effect (Table~\ref{tab:table}), it shares the counterintuitive non-monotonic temperature dependence of the time to reach the final state---a high-temperature bath for the Mpemba effect and a gas phase in equilibrium with a low-temperature bath for the Leidenfrost effect. Both effects also involve anomalous heating. Finally, the underlying mechanism for the Leidenfrost effect is well understood, whereas the experimental evidence for and explanations of the inverse Mpemba effect are new.

\onecolumngrid

\vspace{2em}

\vspace{3em}

\begin{table}[htb]
\begin{ruledtabular}
\begin{tabular}{lp{7cm}p{6cm} }

{} & \textit{Leidenfrost} & \textit{Inverse Mpemba} \\ 
\colrule
Initial state & nonequilibrium state at a lower chemical potential & equilibrium state at a lower temperature\\

Reservoir & temperature and chemical potential reservoirs &  temperature reservoir\\
  
Initial temperature & $T_\text{b}$ (reservoir temperature) &  $T_0 \; (< T_\text{b})$ \\

Final temperature & $T_\text{b}$ &  $T_\text{b}$ \\

Mechanism & local heating of droplets & relaxation to equilibrium \\

Final state & equilibrium state at a higher chemical potential &  equilibrium state at a higher temperature\\

\end{tabular}
\caption{\label{tab:table} Comparison between the Leidenfrost and inverse Mpemba effects.}
\end{ruledtabular}
\end{table}

\end{document}